\documentclass[aps,pre,twocolumn,groupedaddress]{revtex4}
\usepackage{graphics}
\usepackage{bm} 
\usepackage{amsmath}

\begin{document}

\title{Construction of bipartite and unipartite weighted networks from collections of journal papers}

\author{Steven A. Morris}
\email[]{samorri@okstate.edu}
\author{Gary G. Yen}
\email[]{gyen@okstate.edu}
\affiliation{
Oklahoma State University\\
Electrical and Computer Engineering\\
Stillwater, OK 74078, USA }

\date{\today}

\begin{abstract}
This work presents a model that allows the study of research
specialties through the manifestations of the specialty's social and
epistemological processes in a collection of journal papers.
Collections of papers are modeled as coupled bipartite networks
interlinking 7 types of entities. Matrix-based link weight functions
are introduced to calculate weighted bipartite networks and weighted
unipartite co-occurrence networks in the collection of papers. These
weight calculation methods, when used in conjunction with unweighted
bipartite growth models, produce simple growth models for weighted
networks in collections of papers.
\end{abstract}


\maketitle

\section{Introduction\label{introsec}}
A collection of journal papers is a database of papers that
comprehensively samples the journal literature of a scientific
specialty. As such, the social and epistemological processes of the
specialty are manifested in the complex network of linkages among
entities within the collection of papers. These manifestations are
studied by bibliometricians and subject matter experts to assess the
state of research in a specialty, and such studies are used to
advise managers and policy makers in both government and industry to
facilitate research management.

It is important to develop both complex network models and network
analysis tools that can be applied to collections of papers. Such
tools must be used for the problem of predicting how the underlying
processes of a research specialty are manifested in a collection of
papers, and more importantly, to perform the inverse problem of
modeling research specialty processes from their manifestations in
collections of papers. Examples of useful information about research
specialties to be extracted from collections of papers include: 1)
identifying social structures such as research teams, groups of
experts, and leaders of 'schools of thought', 2) identifying
knowledge structure, such as research subtopics, base knowledge, and
exemplars, and  3) identifying temporal trends and events such as
discoveries, emergence of new specialties and research teams,
knowledge accretion, and creation and obsolescence of concepts and
exemplars.

This paper introduces a structural model of coupled networks in
collections of journal papers and proposes a construction method for
bipartite and unipartite weighted networks from such collections.
The methods presented here constitute an important step in the
effort to apply the developing science of  complex networks theory
to collections of papers and eventually to the study of scientific
specialties as complex social networks and knowledge networks.

As complex networks, collections of papers have three distinguishing
characteristics: 1) they are formed from coupled networks of many
different types of entities, e.g., papers, references, authors, 2)
both unipartite and bipartite networks in collections of papers are
best expressed as weighted networks, where strength of linkage
between pairs of entities is expressed as a positive real link
weight, and 3) collections of papers are best represented as
collections of bipartite networks.

To date, the phenomenon of coupled networks has received little
attention in the physics literature. Zheng and Ergun \cite{zheng03}
model the simultaneous growth of two loosely coupled sections of a
unipartite network and show conditions for power-law link
distributions in the crosslinks between network sections. Borner,
\emph{et al}, model the simultaneous growth of citation networks and
author collaboration networks by modeling behavior of authors
\cite{borner04}.

In contrast to the paucity of research on coupled networks, recently
a great deal of study has been focused on weighted networks. Yook,
\emph{et al} \cite{yook01}, originally investigated growing weighted
networks using preferential attachment rules and random attachment
rules. Newman \cite{newman04analysis} showed that weighted networks
could be expressed as multigraphs, and explained how this treatment
allows generalization of many analysis techniques of unweighted
networks to weighted networks.  Barrat, \emph{et al}
\cite{barrat04c}, studied a large weighted author collaboration
network, and the weighted world airline network, and showed that
these networks have differences in correlations of node degrees to
strength and clustering. Other studies focus on the statistical
properties of weighted networks \cite{barrat04, bianconi04,
almaas04, jezewski04}, transport models of weighted networks
\cite{bagler04, goh05, goh04}, or growth models of weighted networks
\cite{barthelemy05, barrat04b, dorogovtsev04, antal05, fu04}. Fan,
\emph{et al} \cite{fan04},and Li, \emph{et al} \cite{fan04a},
gathered a collection of papers on the specialty of econophysics,
and studied a weighted unipartite collaboration network of authors
from that collection.

On the topic of bipartite networks, recently several papers have
reported on structural models and growth models. Ergun
\cite{Ergun02} models the human sexual contact network as a
bipartite graph, with growth having preferential attachment rules
similar to a Yule process. Ramasco, \emph{et al}, present a
bipartite Yule model for paper to author networks \cite{ramasco04}.
Guillaueme and Latapy \cite{guillaume04a} also present a bipartite
Yule model and propose a method of deriving a bipartite expression
of any unipartite network.  Morris \cite{morris05a} proposes the use
of general bipartite Yule processes for entity-type pairs in
collections of journal papers, and gives examples for paper to
reference networks and paper to author networks. Morris
\cite{morris04a} also gives a detailed analysis of a bipartite Yule
model for paper to reference networks that models heavily cited
exemplar references in emerging specialties. Goldstein, \emph{et
al}, \cite{goldstein04group} and Morris, \emph{et al},
\cite{morris04b} propose bipartite Yule models for paper to author
networks that model the success-breeds-success phenomenon for teams
of authors.

\begin{figure}
\resizebox{0.45\textwidth}{!}{%
\includegraphics{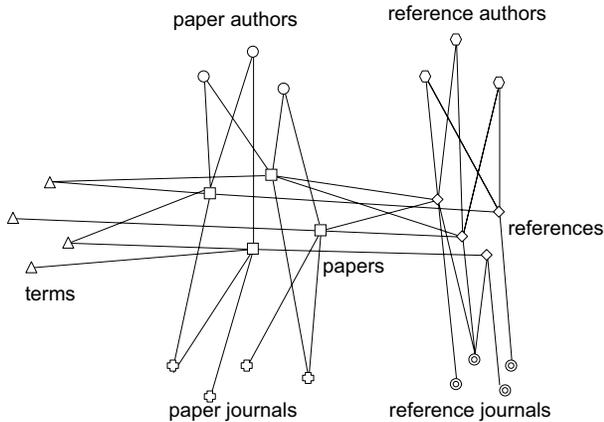}}%
\caption{Diagram showing a collection of papers as a series of
coupled bipartite networks.\label{coupled}}
\end{figure}

As shown in Figure \ref{coupled}, a collection of journal papers
constitutes a series of coupled bipartite networks. As diagrammed in
the figure, a collection of papers contains 6 direct bipartite
networks: 1) papers to paper authors, 2) papers to references, 3)
papers to paper journals, 4) papers to terms, 5) references to
reference authors, and 6) references to reference journals.
Additionally, there are 15 indirect bipartite networks in
collections of papers as defined by the diagram. Examples of
interesting indirect networks are paper authors to reference
authors, and paper journals to reference journals networks, which
can be used for author co-citation analysis \cite{white81} and
journal co-citation analysis \cite{mccain91} respectively.

This paper introduces a formal matrix-based treatment of coupled
bipartite structures in collections of papers.  This treatment is
used to calculate the weights of indirect bipartite networks and is
extended to calculation of weights of unipartite co-occurrence
networks in the collection. For example, the proposed method can be
used to calculate the weights of a bipartite paper author to
reference network, or, it can be used to find weights of the
unipartite co-occurrrence network of authors that link to common
papers (a co-authorship network).

The proposed matrix-based technique is similar to multi-port
analysis using ABCD parameters in electrical networks
\cite{chirlian69}.  The method is also very similar to methods used
in multi-layer neural networks \cite{hagan96}.

In conjunction with simple bipartite Yule growth models
\cite{morris05a}, the proposed weight calculation method produces
 simple models of weighted network growth, growing as it does from
unweighted direct links that occur as papers are added to the
collection.

\section{Collections of journal papers}
\subsection{Research specialties}

A \textit{research specialty}  is a self-organized social
organization whose members tend to study a common research topic,
attend the same conferences, publish in the same journals, cite each
other's work, and belong to the same social networks that are known
as \textit{invisible colleges} \cite{crane72}. Thomas Kuhn, the
pioneer of the study of research processes, considered specialties
to be quite small, "100 members, sometimes considerably less"
\cite{kuhn70}.

The processes that drive research specialties are twofold: 1) social
processes of research teams, communication networks, and
collaboration, and 2) epistemological processes of the discovery,
emergence, accretion, and obsolescence of knowledge. As described by
Kuhn, the distinguishing feature of a specialty is its
\textit{paradigm}, which is the researchers' "way of thinking" about
their problem: models, analytical techniques, validation standards
and so forth. Progress in a specialty is characterized by long and
stable periods of \textit{puzzle-solving} within the specialty's
paradigm, punctuated by discoveries that accompany the overthrow
and/or creation of new paradigms \cite{kuhn70}. This characteristic
of specialties is similar to \textit{punctuated equilibria}
phenomena \cite{eldredge72} that characterize self-organizing
systems \cite{bak96}.

Specialties create their own \emph{literature}, i.e., a body of
journal papers and books that broadly focus on the specialty's
research topic. We define a \textit{collection of papers} as a list
of journal papers that constitutes a comprehensive sample of a
specialty's journal literature. As a working definition, define a
collection of papers as a database of records, one record per paper,
that contains information about the individual papers in such a
list.

Although the range of size of such collections is large, the size of
such collections is much smaller than the immense databases of
papers that are often studied in the physics literature. Morris
\cite{morris04a}, using back-of-envelope style approximations,
suggests that collections of papers should range from as few as 100
papers to as many as 5000 papers. Huge heterogeneous datasets, such
as the SPIRES database \cite{redner98}, 20 years of PNAS papers
\cite{borner04}, 100 years of Physical Review journals
\cite{redner04}, or all the chemistry publications of the
Netherlands \cite{vanraan01}, are not collections of papers as
defined here, because they all sample more than one specialty's
literature. Despite this conceptual constraint, the weight
calculation method proposed here can still be applied to such huge
collections.

\subsection{Definition of collections of journal papers} For
discussion in this paper, a collection of journal papers is a
database where each record corresponds to a journal paper. For each
paper, its associated authors, cited references, journal, index
terms and publication year are listed. Furthermore, for each
reference, a reference author, reference journal, and reference year
are listed. As defined here, collections of papers are constructed
to comprehensively sample the literature of a scientific specialty.
For our purposes, collections of papers are typically downloaded
from the Science Citation Index using Thompson/ISI's Web of Science
product \footnote{http://www.isinet.com}. Queries and seed
references are used to gather topic specific collections that cover
a specialty.  The records for these papers are typically collected
into text files using a tagged file format and downloaded for
analysis. For the purpose of demonstrating the concepts proposed in
this paper, a fictitious collection of four papers is given in the
Appendix that covers the fictional specialty of
\textit{improbability generation}. (Apologies to humor author
Douglas Adams.) This example collection is provided to allow readers
to understand the extraction of entities and links from the source
data of the collection. For illustrative purposes the entities in
this example are more densely linked than would normally be found in
such a small collection of papers.

A collection of papers can be considered as a network of
\textit{bibliographic entities} of various \textit{entity-types}
\cite{morris04crossmaps}. Bibliographic entities may correspond to
\textit{physical entities} in the real world, and more than one
bibliographic entity may correspond to the same physical entity. For
example, a paper and a reference in a collection of papers may both
correspond to the same physical paper in the real world.

It is common in studies of networks in journal literature to match
references to papers to build a model of "papers citing papers",
usually referred to as a \textit{citation network} \cite{albert02}.
There are both methodological and theoretical reasons to avoid this
type of treatment: 1) on one hand, a collection of papers typically
has 20 times more references than papers, making such citation
network models grossly incomplete because unmatched papers and
references (including references corresponding to books), have
unknown incoming and outgoing links, 2) the second problem is that
references, especially highly cited references, can be considered as
\textit{concept symbols} \cite{morris04a, small78}, and therefore
should be considered as separate entity-types from papers, which
merely represent undifferentiated research reports. Figuratively, it
is inappropriate to use an "apples-citing-apples" model when the
actual network is "apples-citing-oranges." Further discussion of
citation networks is outside the scope of this paper.

For our proposed structural model of collections of journal papers
presented in this paper, we will limit our discussion to a model
comprised of 7 entity-types: 1) papers, 2) paper authors, 3) paper
journals, 4) index terms, 5) references, 6) reference authors and 7)
reference journals. Index terms are terms supplied by authors or
abstract services to associate with papers for search and
classification purposes. Paper authors are the authors of papers,
while reference authors are the authors associated with references.
Paper journals are the journals that papers are published in, while
reference journals are the journals associated with references.
References corresponding to books, films, web pages, and eprint
archive articles have no associated reference journal.

Using the 7 entity-types given in our structural model, Figure
\ref{coupled} illustrates that a collection of journal papers
constitutes a series of coupled bipartite networks. As noted in
Section \ref{introsec}, there are 6 direct bipartite networks and 15
indirect bipartite networks in this structural model. These indirect
bipartite networks are best analyzed as weighted networks and those
weights can be calculated from the paths of direct links that
connect entities in the two partitions of interest.

Note the fictitious collection of papers in the Appendix. The source
file for this collection, which consists of 4 papers, is listed in
ISI tagged file format. See footnote \footnote{A set of MATLAB
routines that can extract several types of bipartite networks from
ISI tagged files is available from the authors. Please contact one
of the authors for further information}. The extracted entities for
this collection consists of 4 papers, 3 paper authors, 4 paper
journals, 7 index terms, 10 references, 6 reference authors, and 7
reference journals. These entities and their corresponding index
numbers are listed in the Appendix.

\section{Bipartite networks in collections of journal papers}
\subsection{Dyad definitions} In a dyad, the two entities can be: 1) \textit{like entities}, that is,
entities of the same entity-type, or 2) \textit{unlike entities},
that is, entities of different entity-types. \textit{Direct links}
are defined as direct associations. A paper has direct links to its
authors (paper authors), its associated index terms, the references
the paper cites, and the journal the paper was published in.  A
reference is directly linked to the papers that cite it, the author
associated with the reference (reference author), and the journal
that is associated with the reference (reference journal).

\textit{Indirect links} are links between two unlike entities that
occur over a path of two or more direct links. For example, a paper
author is indirectly linked to a reference author if he or she
authors a paper that cites a reference that is associated with that
reference author.

The first entity of interest in a dyad is the \textit{primary
entity} while the other entity is the \textit{secondary entity}.
Designation of primary entity-type and secondary entity-type in
direct and indirect bipartite networks is arbitrary and is assumed
to be based on the interest of the investigator. For
\emph{co-occurrence networks}, the primary and secondary
entity-types are explicitly defined, as will be explained in Section
\ref{cooccursec}. \textit{Co-occurrence links} are between like
primary entities and occur when both entities link to the same
secondary entity.  For example, two papers have a co-occurrence link
when they both cite a common reference, or, in another example, two
paper authors have a co-occurrence link if they coauthor a paper. In
co-occurrence links the like entities of the dyad are primary
entities, while the unlike entities to which they co-link are the
secondary entities.

\subsection{Dyad identifier notation} Table \ref{deftab} lists the
conventions used here to denote entity-type variables within a
collection of papers.  The variables $x_1$, $x_2$, and so forth will
be used to denote unspecified entity-types. \textit{Dyad notation}
is used to specify dyad types in the collection of papers. The
symbols of primary and secondary entity-types associated with dyads
are separated by a comma and placed between square brackets, e.g.,
$[x_1,x_2]$, where $x_1$ denotes the primary entity-type, and $x_2$
denotes the secondary entity-type. This notation will be referred to
as the \textit{dyad identifier}, and will be used as a suffix to
variables to specify the entity-types of interest. However, the dyad
identifier will be dropped to reduce clutter in the notation when
the primary and secondary entity-types are obvious from context.
Some examples of the use of dyad identifiers:
\begin{itemize}
\item $\mathbf{O}[p,r]$ denotes an occurrence matrix listing the links of
papers, the primary entity-type, to references, the secondary
entity-type.
\item$\mathbf{C}[ap,p]$  denotes the co-occurrence matrix listing the
co-authorship counts of pairs of paper authors, the primary
entity-type, in papers, the secondary entity-type.
\end{itemize}

\begin{table}
\caption{Variable conventions used for entities in collections of
papers.\label{deftab}}
\begin{tabular}{|p{.2\textwidth}l|p{.2\textwidth}l|} \hline

$p$: paper & $r$: reference\\
$ap$: paper author & $ar$: reference author\\
$jp$: paper journal & $jr$: reference journal\\
$yp$: paper year & $yr$: reference year\\
$t$: term & \\
$x_i$: unspecified entity  & \\ \hline
\multicolumn{2}{|p{.45\textwidth}|}{Prefix '$n$' to any entity
variable to denote the number of entities in the
collection of that entity-type, e.g., $np$ denotes the number of papers in the collection}  \\
\hline
\end{tabular}
\end{table}

\subsection{Bipartite networks} 
\begin{figure}
\resizebox{0.25\textwidth}{!}{%
\includegraphics{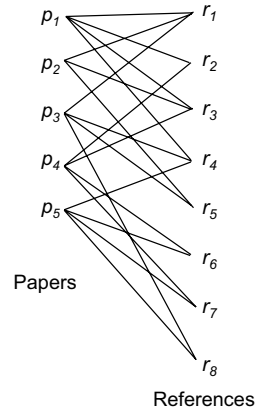}}%
\caption{A collection of papers and references as a bipartite
network. References are linked to papers in which they are
cited.\label{f3}}
\end{figure}

Bipartite
networks are comprised of two distinct partitions of nodes, where
all links in the network are from entities in the first partition to
entities in the second partition. For our purposes, the first
partition exclusively holds entities of some entity-type, while the
other partition exclusively holds entities of some other
entity-type. As an example, Figure \ref{f3} shows a diagram of a
bipartite network of a partition of papers linked to a partition of
references.  Note that links only occur between papers and
references and that there are no links between pairs of papers or
pairs of references.

Assume the diagrammatic convention as shown in Figure \ref{f4}, that
entities of $x_1$, the primary entity-type, are the entities in the
group on the left and the entities of $x_2$, the secondary
entity-type, are the entities in the group to the right.  There are
$nx_1$ primary entities and $nx_2$ secondary entities.  The strength
of the link between $x_1$ entity $i$ and $x_2$ entity $j$ is the
link weight, $o_{ij}[x_1,x_2]$.

\subsection{Occurrence matrices} Mathematically, the links in a
bipartite network are described by a rectangular adjacency matrix,
which we'll define as an \textit{occurrence matrix}. This is an
$nx_1$ by $nx_2$ matrix that lists all the link weights between the
entities of the two partitions:
\begin{equation}
\mathbf{O}[x_1,x_2]=
 \left[
\begin{array}{cccc}
o_{11} & o_{12} & \dots & o_{1nx_2} \\
o_{21} & \ddots & & \vdots\\
\vdots & & \ddots & \vdots \\
o_{nx_11} & \dots & \dots & o_{nx_1nx_2} \\
\end{array}
\right]
\end{equation}

\begin{figure}
\resizebox{0.25\textwidth}{!}{%
\includegraphics{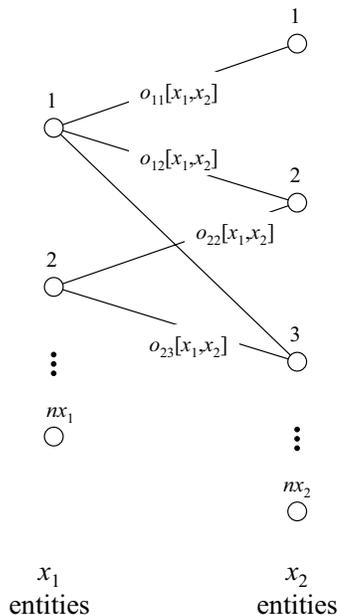}}%
\caption{Diagram of a general bipartite network and conventions for
labeling link weights in the occurrence matrix of the network.
\label{f4}}
\end{figure}

Figure \ref{f4} shows how the links in a bipartite network
correspond to elements in its occurrence matrix. There is a
bipartite network for every possible pair of entity-types in the
collection of papers. Occurrence matrices for entity-type pairs with
direct relations are derived directly from the tables in the
collection's database. For the example collection of papers
discussed in this paper, the occurrence matrices for the 6 direct
bipartite networks in the collection are given in the Appendix.
Occurrence matrices for entity-type pairs with indirect links are
calculated by cascading bipartite networks of direct links, as will
be shown later.

Note the following property of occurrence matrices:

\begin{equation}
\mathbf{O}[x_1,x_2]=\mathbf{O}[x_2,x_1]^T \label{eq26}
\end{equation}

Using dyad identifier notation, exchanging the variables is
equivalent to transposing the occurrence matrix.

\subsection{Coupled and cascaded bipartite
networks\label{coupledsec}} 
\begin{figure}
\resizebox{0.45\textwidth}{!}{%
\includegraphics{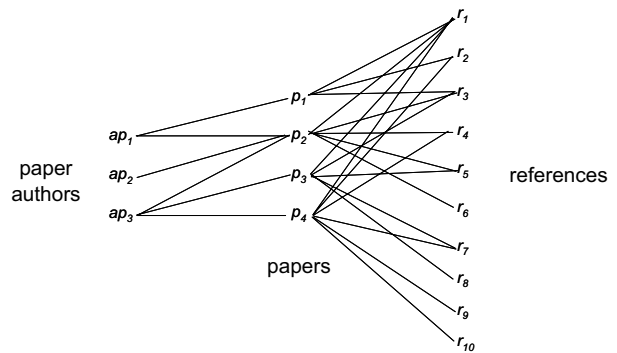}}%
\caption{An example of coupled bipartite networks. A paper
author-paper network is coupled to a paper-reference network through
common papers. This example is taken from the example collection in
the Appendix.\label{coupled1}}
\end{figure}

\begin{figure}
\resizebox{0.45\textwidth}{!}{%
\includegraphics{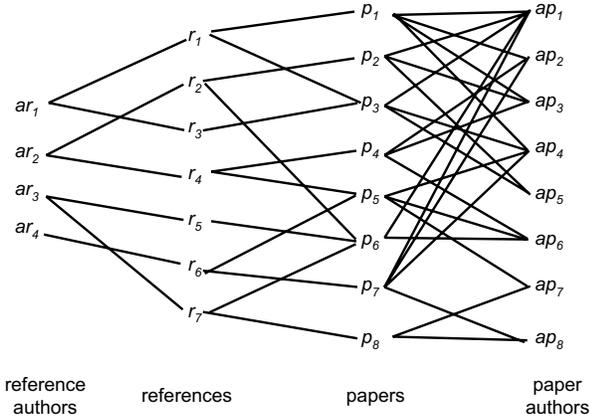}}%
\caption{An example of a cascade of bipartite networks. A reference
author to reference network is coupled to a reference to paper
network that is, in turn, coupled to a paper to paper author
network.\label{cascade}}
\end{figure}

\textit{Coupled bipartite networks} are
pairs of bipartite networks that share a common partition. Figure
\ref{coupled1} shows an author to paper network coupled to a paper
to reference network through common papers using the example
collection of papers in the Appendix.  \textit{Cascaded bipartite
networks} are comprised of a series of two or more coupled bipartite
networks. Figure \ref{cascade} shows an example of such a cascade,
where a reference author to reference network is coupled to a
reference to paper network that is in turn coupled to a paper to
paper author network. We define the extreme left and right
partitions as the \textit{outer partitions} and all other partitions
as the \textit{inner partitions}.

Assume that we are interested in describing the links between two
different types of entities as a weighted bipartite network. We
first find a cascade of networks where the two entity-types of
interest are the outer partitions. Then it is necessary to apply
some algorithm that meaningfully reduces the indirect links between
pairs of opposite outer entities as weights in a bipartite network
joining those outer entities. Intuitively, we want pairs of outer
entities that have many indirect links through the inner partitions
to have more weight than those pairs of outer entities with few or
no connecting links.

For example, suppose that we wish to find a weighted bipartite
network between reference authors and paper authors for the purpose
of conducting author co-citation analysis \cite{white81}. We can
find a cascade of bipartite networks as shown in Figure
\ref{cascade}, where reference authors are linked to their
references, the references are linked to the papers that cite them,
and those papers are linked to the paper authors that authored them.
The weights of a bipartite network of reference authors to paper
authors are found by finding the indirect links between each
reference author and paper author through references and papers, and
applying an algorithm that produces a weight from those identified
indirect links. The more indirect links between a reference author
and a paper author, the more weight should be assigned to the link
between them in the resulting bipartite network.

\section{Algorithm for construction of weighted bipartite networks\label{constructsec}}
\subsection{Reducing a cascade of bipartite networks to a single
weighted bipartite network}

Given a cascade of bipartite networks with occurrence matrices
$\mathbf{O}[x_1,x_2]$, $\mathbf{O}[x_2,x_3],\dots,
\mathbf{O}[x_{n-1},x_n]$, this cascade can be reduced to a single
bipartite network  with occurrence matrix $\mathbf{O}[x_1,x_n]$
listing the link weights between the $x_1$ entities and the $x_n$
entities in the network. The proposed weight algorithm is iterative
and works by sequentially reducing two adjacent networks to a single
network, then reducing that weighted network and its adjacent
network.  This process continues until only a single bipartite
network remains.

The algorithm is based on using a generalized form of matrix
arithmetic. Given a pair of opposite outer entities, the algorithm
finds all unique paths from the left outer entity to the right outer
entity, and assigns a weight to each of those paths. The weights of
these parallel paths are then combined to calculate the weight of
the link between the two entities.

\subsection{Reducing adjacent coupled bipartite networks to a single weighted bipartite network}

\begin{figure}
\resizebox{0.4\textwidth}{!}{%
\includegraphics{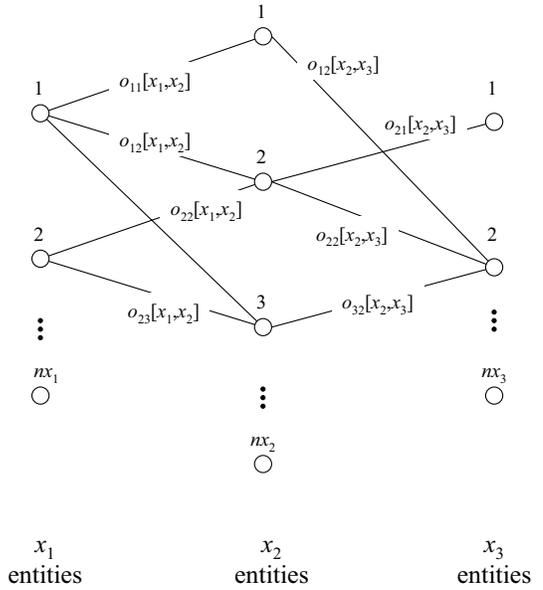}}%
\caption{Diagram of adjacent bipartite networks and conventions for
naming entities and links.\label{f5}}
\end{figure}

\begin{figure}
\resizebox{0.45\textwidth}{!}{%
\includegraphics{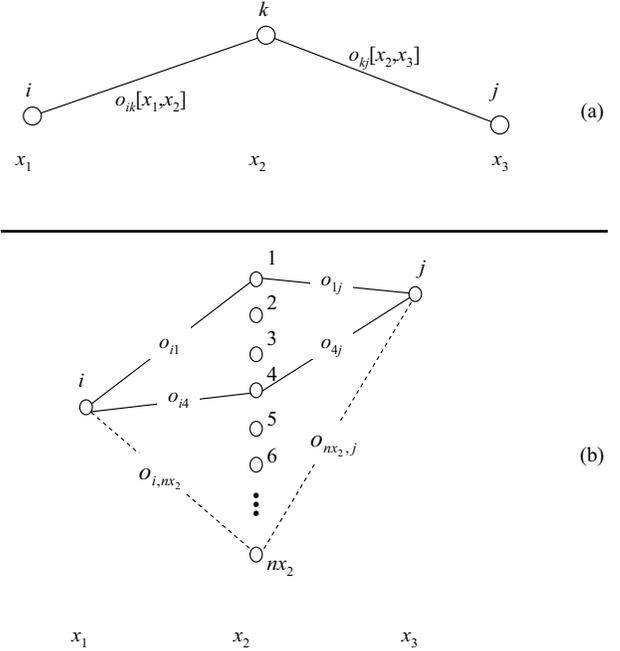}}%
\caption{ a) Example path between $x_1$ entity $i$ and $x_3$ entity
$j$ through $x_2$ entity $k$. b) Shows $nx_2$ possible paths between
$x_1$ entity $i$ and $x_3$ entity $j$ through $x_2$
entities.\label{f6}}
\end{figure}

Consider a pair of coupled bipartite networks, with entity-types
$x_1$, $x_2$, and $x_3$, as shown in Figure \ref{f5}. Occurrence
matrices $\mathbf{O}[x_1,x_2]$ and $\mathbf{O}[x_2,x_3]$ enumerate
the links in the two bipartite networks in this figure. Each link in
the figure is labeled with its corresponding occurrence matrix
element. There are $nx_1$, $nx_2$, and $nx_3$ entities of the
entity-types $x_1$, $x_2$, and $x_3$ respectively. A pair of links
that connects an $x_1$ entity to an $x_3$ entity is defined as a
\emph{path}. Figure \ref{f6}, part (a) shows a path from $x_1$
entity $i$ to $x_3$ entity $j$, connected through $x_2$ entity $k$
by links $o_{ik}[x_1,x_2]$ and $o_{kj}[x_2,x_3]$. There are $nx_2$
possible paths from $x_1$ entity $i$ to $x_3$ entity $j$ as shown in
Figure \ref{f6} part (b).

The \textit{path weight} associated with a path is calculated from
the weights of the path's two links using a \textit{path weight
function}:
\begin{equation}
p_{ij}(k)=f_2(o_{ik}[x_1,x_2], o_{kj}[x_2,x_3])\label{eq1},
\end{equation}
where $f_2$ is the path weight function, to be defined later.  The
resulting link weight from $x_1$ entity $i$ to $x_3$ entity $j$ is
calculated from the path weights of all possible paths between those
two entities using a \textit{path combining function}:
\begin{equation}
o_{ij}[x_1,x_3]=f_1\Big(p_{ij}(1), p_{ij}(2), \dots
p_{ij}(nx_2)\Big), \label{eq2}
\end{equation}
where $f_1$ is the path combining function, to be defined later.
Substituting Equation (\ref{eq1}) into Equation (\ref{eq2}) gives
the \textit{link weight function} which defines the rules for
calculating link weights of cascaded bipartite networks:
\begin{multline}
o_{ij}[x_1,x_3]=\\
f_1 \Big( f_2(o_{i1},o_{1j}), f_2(o_{i2},o_{2j}) ,\dots,
f_2(o_{i\,nx_2},o_{nx_2\, j})\Big). \label{eq3}
\end{multline}
\begin{figure}
\resizebox{0.45\textwidth}{!}{%
\includegraphics{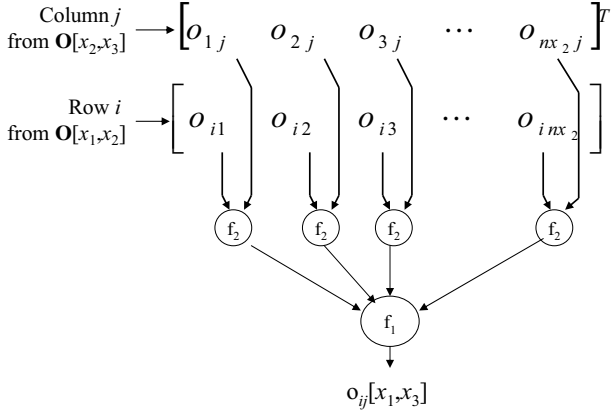}}%
\caption{Diagram illustrating vector operation of the link weight
function.\label{f7}}
\end{figure}

The link weight function of Equation \ref{eq3} is a matrix function
that is used to compute all the $nx_1$ times $nx_3$ possible weights
of the occurrence matrix $\mathbf{O}[x_1,x_3]$ according to the
rules for weight computation given by $f_1$ and $f_2$. Consider
Figure \ref{f7}  which illustrates how the link weight function uses
row $i$ of $\mathbf{O}[x_1,x_2]$ and column $j$ of
$\mathbf{O}[x_2,x_3]$ to produce element $o_{ij}$ of matrix
$\mathbf{O}[x_1,x_3]$. As shown, the function $f_2$ is applied to
matching elements of the row vector and column vector to produce
$nx_2$ scalar results. The function $f_1$ operates on all these
$nx_2$ results to produce the final scalar result $o_{ij}[x_1,x_3]$.

The concepts of 1) bipartite networks of entities, 2) cascaded
bipartite networks, and 3) link weight functions, provide a
systematic means of finding multiple indirect links between outer
entities in cascades of bipartite networks, and combining those
multiple links as a weight in a bipartite network between the outer
entities. The choice of path weight function and path combining
function is generally driven by the application.  In the case of
cascades of unweighted bipartite networks, matrix multiplication
makes a good link weight function because it yields weights that are
equal to occurrence counts. For example, for a paper to reference
network coupled to a reference to reference author network, matrix
multiplication as a link weight function will produce weights,
$o_{ij}[p,ar]$, that are the the number of times paper $i$ cites
reference author $j$.

In other situations, however, other link weight functions are more
appropriate. For example, when reducing cascades of weighted
bipartite networks, it is necessary to consider how to compute path
weights from the two links in a path. Suppose we have a weighted
bipartite network of \emph{linguistic terms} to papers in a
collection of papers. The weights, $o_{ij}[t,p]$, in this network
are the number of times term $i$ appears in the body of paper $j$.
Now assume this matrix is coupled to a paper to reference author
network, and that there is a path from term $i$ to reference author
$j$ that corresponds to 10 occurrences of term $i$ in paper $k$,
which cites reference author $j$ 2 times. If we use multiplication
as the path weight function, then this yields $10 \times 2 = 20 $
for the path weight.  This has no meaning as an occurrence count
between term $i$ and reference author $j$. In this case we may want
to simply use a link weight equal to the number of times reference
author $j$ is cited by paper $k$, or use a link weight equal to the
minimum of the number of times paper $k$ cites reference author $j$
and the number of times term $i$ occurs in paper $k$.  We can also
express the two links in the path as electrical conductances and
calculate the path weight as the resulting conductance of those two
conductances in series.

The next three subsections will describe three link weight
functions: 1) matrix multiplication, appropriate for cascades of
unweighted networks, 2) the overlap function, appropriate for
cascades of weighted occurrence networks, and 3) the inverse
Minkowski function, used to compute paths weights as similar to
conductances in series.

\subsection{Link weight function using matrix multiplication}

For applications where at least one of the matrix arguments is
binary, matrix multiplication is often used as the link weight
function because it directly yields  weights that are simple
occurrence and co-occurrence counts in the resulting reduced
bipartite matrix.

If the path weight function $f_2$ is defined as a product:
\begin{equation}
f_2\biggl(o_{ik}[x_1,x_2],o_{kj}[x_2,x_3]\biggr)=o_{ik}[x_1,x_2]\cdot
o_{kj}[x_2,x_3] \label{eq5}
\end{equation}
and the path combining function $f_1$ is a summation:
\begin{multline}
f_1\Bigl(f_2\bigl(o_{i1}[x_1,x_2],o_{1j}[x_2,x_3]\bigr),\dots, \\
f_2\bigl(o_{i\,nx_2}[x_1,x_2],o_{nx_2\,j}[x_2,x_3]\bigr)\Bigl)  \\
  = \sum_{k=1}^{nx_2}f_2\bigl(o_{ik}[x_1,x_2],o_{kj}[x_2,x_3]\bigr).\label{eq6}
\end{multline}
Then the link weight function is simply standard matrix
multiplication:
\begin{equation}
o_{ij}[x_1,x_3]=\sum_{k=1}^{nx_2} o_{ik}[x_1,x_2]\cdot
o_{kj}[x_2,x_3]. \label{eq7}
\end{equation}

As an example, assume that $x_1$, $x_2$, and $x_3$ are paper
authors, papers and references respectively, taken from the example
collection of papers in the Appendix.  The binary matrix
$\mathbf{O}[ap,p]$, the transpose of $\mathbf{O}[p,ap]$, Equation
(\ref{opap}), lists the links of the individual paper authors to
each paper, while the binary matrix $\mathbf{O}[p,r]$, Equation
(\ref{opr}), lists the links of individual papers with each
reference. Using matrix multiplication:
\begin{equation}
\mathbf{O}[ap,r]=\mathbf{O}[ap,p]\cdot \mathbf{O}[p,r]. \label{eq10}
\end{equation}
This yields:
\begin{eqnarray}
\mathbf{O}[ap,r] &=& \left[
\begin{array}{cccc}
 1 & 1 & 0 & 0 \\
 0 & 1 & 0 & 0 \\
 0 & 1 & 1 & 1 \\
\end{array}
\right]
 \left[\begin{array}{cccccccccc}
 1 & 1 & 1 & 0 & 0 & 0 & 0 & 0 & 0 & 0 \\
 1 & 0 & 1 & 1 & 1 & 1 & 0 & 0 & 0 & 0 \\
 1 & 0 & 1 & 0 & 1 & 0 & 1 & 1 & 0 & 0 \\
 1 & 1 & 0 & 1 & 0 & 0 & 1 & 0 & 1 & 1 \\
\end{array}\right] \nonumber \\
\nonumber \\
&=&  \left[
\begin{array}{cccccccccc}
 2 & 1 & 2 & 1 & 1 & 1 & 0 & 0 & 0 & 0 \\
 1 & 0 & 1 & 1 & 1 & 1 & 0 & 0 & 0 & 0 \\
 3 & 1 & 2 & 2 & 2 & 1 & 2 & 1 & 1 & 1 \\
\end{array}
\right]. \label{eq11}
\end{eqnarray}
This is a matrix, $\mathbf{O}[ap,r]$, in which weight,
$o_{ij}[ap,r]$, is the number of times that paper author $i$ cites
reference $j$.

Suppose we wish to find the paper author to reference author
occurrence matrix of the example collection of papers in the
Appendix. Consulting Figure \ref{coupled}, the direct links from
paper authors to reference authors go from paper author to paper to
reference to reference author. Calculation of the occurrence matrix,
$\mathbf{O}[ap,ar]$, from paper author to reference author is
performed by the matrix multiplication:
\begin{equation}
\resizebox{0.35\textwidth}{!}{%
\includegraphics{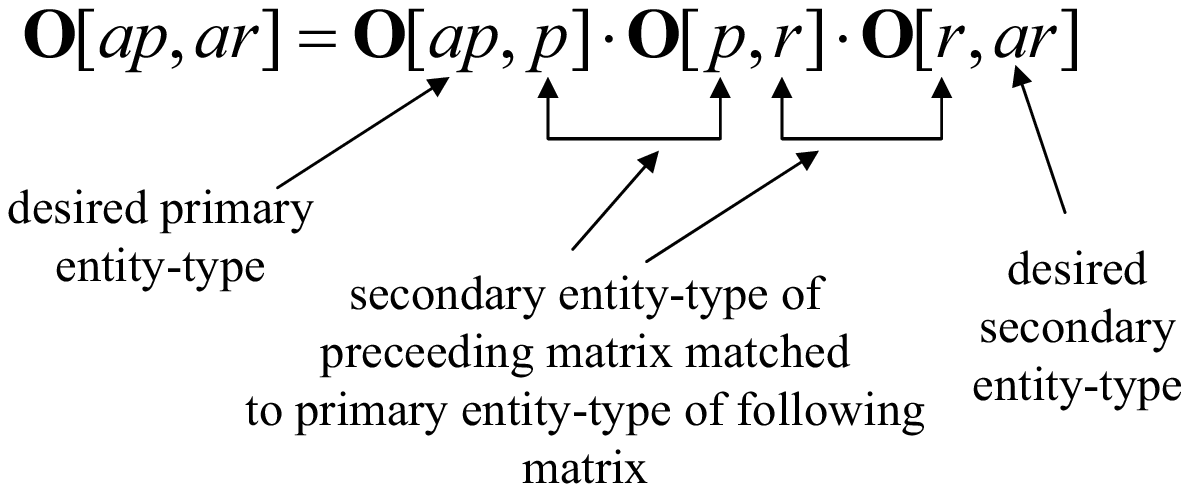}}%
.\end{equation}

Using the example paper collection in the Appendix, first find the
paper author to reference matrix by multiplying the paper author to
paper matrix and the paper to reference matrix. This was done in
Equation (\ref{eq11}). Then multiply the paper author to reference
matrix with the reference to reference author matrix:
\begin{eqnarray}
\mathbf{O}[ap,ar]=\mathbf{O}[ap,r]\cdot\mathbf{O}[r,ar]&=& \nonumber \\
\nonumber \\
 \left[
 \begin{array}{cccccccccc}
 2 & 1 & 2 & 1 & 1 & 1 & 0 & 0 & 0 & 0 \\
 1 & 0 & 1 & 1 & 1 & 1 & 0 & 0 & 0 & 0 \\
 3 & 1 & 2 & 2 & 2 & 1 & 2 & 1 & 1 & 1 \\
 \end{array}\right]
 \left[\begin{array}{cccccc}
 1 & 0 & 0 & 0 & 0 & 0 \\
 0 & 1 & 0 & 0 & 0 & 0 \\
 0 & 1 & 0 & 0 & 0 & 0 \\
 0 & 0 & 1 & 0 & 0 & 0 \\
 0 & 0 & 1 & 0 & 0 & 0 \\
 0 & 0 & 0 & 1 & 0 & 0 \\
 0 & 0 & 0 & 1 & 0 & 0 \\
 0 & 0 & 0 & 1 & 0 & 0 \\
 0 & 0 & 0 & 0 & 1 & 0 \\
 0 & 0 & 0 & 0 & 0 & 1 \\
\end{array}\right] &=& \nonumber \\
\nonumber \\
 = \left[\begin{array}{cccccc}
 2 & 3 & 2 & 1 & 0 & 0  \\
 1 & 1 & 2 & 1 & 0 & 0  \\
 3 & 3 & 4 & 4 & 1 & 1  \\
\end{array}\right]\label{eq32}
.\end{eqnarray}

The result in Equation (\ref{eq32}) gives the desired occurrence
matrix of paper authors to reference authors for the example. In
this matrix, the weight $o_{ij}[ap,ar]$ is the number of times that
paper author $i$ cites reference author $j$.

\subsection{Link weight function using the overlap
function\label{overlapsec}} The overlap function is useful for
calculating weights of links when reducing cascades of weighted
bipartite networks. This is appropriate for calculating bipartite
networks involving linguistic terms, and is also useful for
calculating weights in co-occurrence networks of reference authors
and reference journals.

Think of the two links in a path as conduits, each with a maximum
capacity. The maximum capacity of these two conduits in series is
equal to that of the conduit with the smallest capacity. Considering
this series capacity as the path weight, the path weight function
becomes the minimum of the weights of the two links on the path:
\begin{equation}
f_2=min\Bigl(o_{ik}[x_1,x_2],o_{kj}[x_2,x_3]\Bigr)\label{eq12}
.\end{equation}
Using a path combining function that sums the path
weights:
\begin{equation}
f_1=\sum_{k=1}^{nx_2} f_2\Bigl(o_{ik}[x_1,x_2],o_{kj}[x_2,x_3]\Bigr)
\label{eq13} ,\end{equation} yields the overlap function
\cite{salton89} as the link weight function:
\begin{equation}
f_1=\sum_{k=1}^{nx_2} min\Bigl(o_{ik}[x_1,x_2],o_{kj}[x_2,x_3]\Bigr)
\label{eq14} .\end{equation}
This can be defined as a matrix
operation "OVL":
\begin{equation}
\mathbf{O}[x_1,x_3]=OVL\Bigl(
\mathbf{O}[x_1,x_2],\mathbf{O}[x_2,x_3] \Bigr) \label{eq15}
.\end{equation}
Discussion of the application and characteristics of
this function can be found in \cite{jones87}.

\begin{figure}
\resizebox{0.45\textwidth}{!}{%
\includegraphics{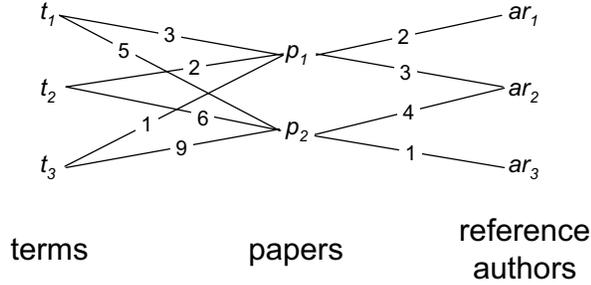}}%
\caption{Example of cascaded bipartite networks with non-binary link
weights.  Terms to paper network cascaded with paper to reference
author network.\label{f9}}
\end{figure}
As an example, assume that $x_1$, $x_2$, and $x_3$ are linguistic
terms, papers and reference authors respectively, as shown in Figure
\ref{f9}. The matrix $\mathbf{O}[t,p]$ lists the occurrence counts
of the individual terms with each paper:
\begin{equation}
\mathbf{O}[t,p]=
 \left[
\begin{array}{cccc}
 3 & 5  \\
 2 & 6  \\
 1 & 9  \\
\end{array}
\right]\label{eq16} ,
\end{equation} and the matrix
$\mathbf{O}[p,ar]$ lists the associations of individual papers with
each reference author:
\begin{equation}
\mathbf{O}[p,ar]=
 \left[
\begin{array}{cccc}
 2 & 3 & 0 \\
 0 & 4 & 1 \\
\end{array}
\right]\label{eq17} .\end{equation} Using the overlap function to
calculate the link weights of $\mathbf{O}[t,ar]$:
\begin{eqnarray}
\mathbf{O}[t,ar]&=&OVL\Bigl(\mathbf{O}[t,p],\mathbf{O}[p,ar]\Bigr)\nonumber \\
\nonumber\\
 \mathbf{O}[t,ar]&=&OVL\left(
 \left[
\begin{array}{cc}
 3 & 5  \\
 2 & 6  \\
 1 & 9  \\
\end{array}\right],
\left[
\begin{array}{ccc}
 2 & 3 & 0 \\
 0 & 4 & 1 \\
\end{array}
\right] \right)
 =\left[\begin{array}{ccc}
 2 & 7 & 1  \\
 2 & 6 & 1 \\
 1 & 5 & 1 \\
\end{array}\right].
\label{eq19}\nonumber \\
\end{eqnarray}

\subsection{Link weight function using the inverse Minkowski
function\label{minkowskisec}}

The \textit{inverse Minkowski function}, an adaptation of the
well-known Minkowski distance metric \cite{cios98}, can be used when
it is desired to model path weights as if the link weights were
electrical conductances in series.   In this case use the inverse
Minkowski metric as the path weight function:

\begin{equation}
f_2=\left[{ \Bigl( {o_{ik}[x_1,x_2]} \Bigr) }^{-p} +
{\Bigl(o_{kj}[x_2,x_3]\Bigr)}^{-p} \right]^{-\frac{1}{p}}
\label{eq20} ,\end{equation} where $p$ ranges from zero to positive
infinity.  Note that, in contrast to the Minkowski metric as
normally expressed, the exponents in the inverse Minkowski metric
are negative. This function will always generate a path weight that
is less than or equal to the smallest link weight in the path,
modeling a situation where indirect links tend to be weaker than
direct links. Using a path combining function that sums the path
weights:
\begin{equation}
f_1=\sum_{k=1}^{nx_2} f_2\Bigl(o_{ik}[x_1,x_2],o_{kj}[x_2,x_3]\Bigr)
\label{eq21}
\end{equation}
yields the final inverse Minkowski link weight function:
\begin{equation}
o_{ij}[x_1,x_3]=\sum_{k=1}^{nx_2}{ \left[{ \Bigl( {o_{ik}[x_1,x_2]}
\Bigr) }^{-p} + {\Bigl(o_{kj}[x_2,x_3]\Bigr)}^{-p}
\right]^{-\frac{1}{p}} }.\label{eq22}
\end{equation}
This can be defined as a matrix operation "$INVMINK$":
\begin{equation}
\mathbf{O}[x_1,x_3]=INVMINK
\Bigl(\mathbf{O}[x_1,x_2],\mathbf{O}[x_2,x_3]\Bigr) \label{eq23}
.\end{equation}

When this function is used with $p=\infty$, Equation (\ref{eq20})
produces the minimum of its arguments and so reverts to Equation
(\ref{eq12}), making the inverse Minkowski link weight function
revert to the overlap link weight function.  When p = 1, then the
path weight function, Equation (\ref{eq20}), becomes:
\begin{equation}
f_2=\left[{ \frac{1}{o_{ik}[x_1,x_2]} + \frac{1}{o_{kj}[x_2,x_3]} }
\right]^{-1} \label{eq24} .\end{equation} This makes the path weight
function produce a value that is twice the harmonic average of the
link weights of the path.  This is equivalent to calculating the
path weight by modeling the link weights as electrical conductances
in series.

The inverse Minkowski path weight function always produces a path
weight that is less than the smallest weight on the path.  This is
appropriate in situations where indirect paths should have less
weight than direct paths, and mathematically expresses a sensed
diffusion, or weakening, of the strength of linkage when linkage is
indirect.

\subsection{Weights in unipartite co-occurrence
networks\label{cooccursec}}

\emph{Co-occurrence networks} are weighted unipartite networks of
like entities where the links between pairs of entities is the count
of the number of common secondary entities that the two primary
entities both link to. For example, in a \textit{bibliographic
coupling network}, the nodes are papers, and the link weights are
the number of common references cited by each pair of papers. A
\textit{co-occurrence matrix} is the adjacency matrix of a
co-occurrence network.  For binary occurrence matrices the
co-occurrence matrix can be found by post multiplying the occurrence
matrix by its transpose. Using Equation (\ref{eq26}):
\begin{equation}
\mathbf{C}[x_1,x_2]=\mathbf{O}[x_1,x_2]\cdot\mathbf{O}[x_2,x_1]\label{eq44}
,\end{equation} where $\mathbf{C}[x_1,x_2]$ is the co-occurrence
matrix listing the number of common associations of pairs of $x_1$
entities with $x_2$ entities. For example, to calculate the
co-occurrence of papers by their links to references using the paper
to reference matrix from the example collection in the Appendix, use
Equation (\ref{opr}):
\begin{eqnarray}
\mathbf{C}[p,r]=\mathbf{O}[p,r]\cdot\mathbf{O}[r,p]&=& \nonumber \\
\nonumber \\
 \left[\begin{array}{cccccccccc}
 1 & 1 & 1 & 0 & 0 & 0 & 0 & 0 & 0 & 0 \\
 1 & 0 & 1 & 1 & 1 & 1 & 0 & 0 & 0 & 0 \\
 1 & 0 & 1 & 0 & 1 & 0 & 1 & 1 & 0 & 0 \\
 1 & 1 & 0 & 1 & 0 & 0 & 1 & 0 & 1 & 1 \\
\end{array}\right]
\left[\begin{array}{cccc}
 1 & 1 & 1 & 1  \\
 1 & 0 & 0 & 1  \\
 1 & 1 & 1 & 0  \\
 0 & 1 & 0 & 1  \\
 0 & 1 & 1 & 0  \\
 0 & 1 & 0 & 0  \\
 0 & 0 & 1 & 1  \\
 0 & 0 & 1 & 0  \\
 0 & 0 & 0 & 1  \\
 0 & 0 & 0 & 1  \\
\end{array}\right] &=& \nonumber \\
\nonumber \\
 = \left[\begin{array}{cccc}
 3 & 2 & 2 & 2  \\
 2 & 5 & 3 & 2  \\
 2 & 3 & 5 & 2  \\
 2 & 2 & 2 & 6  \\
\end{array}\right]
.\label{eq45}
\end{eqnarray}

The diagonal of the co-occurrence matrix $c_{ii}[x_1, x_2]$ lists
the number of links that each $x_1$ has with entities of the $x_2$
entity-type.  For example, in the bibliographic coupling matrix,
$\mathbf{C}[p,r]$, calculated in Equation (\ref{eq45}), the diagonal
lists the number of references each papers cites.

\begin{figure}
\resizebox{0.5\textwidth}{!}{%
\includegraphics{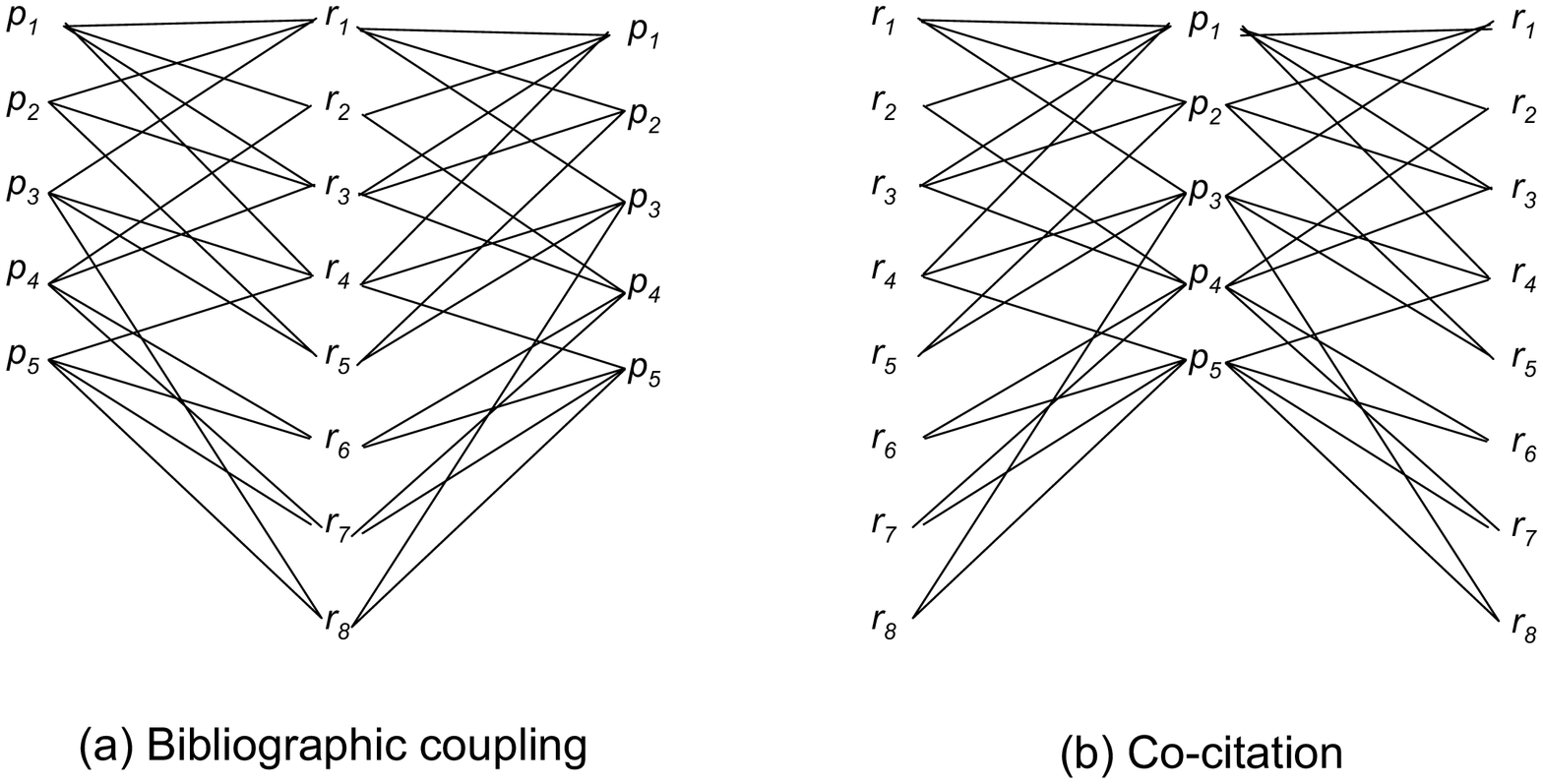}}%
\caption{Mirror of paper to reference bipartite network to calculate
weights in a unipartite co-occurrence network as a cascade of two
bipartite networks. (a) Mirror across references to calculate
bibliographic coupling. (b) Mirror across papers to calculate
co-citation.\label{f13}}
\end{figure}

Computation of co-occurrences can be viewed, similar to the
discussion of Section \ref{coupledsec}, as the calculation of link
weights in a cascade of two bipartite networks.  Given a bipartite
network of two unlike entity-types, mirror the network across the
secondary entity-type partition to obtain a cascade of two networks.
For example, the paper to reference network shown in Figure \ref{f3}
has been mirrored on the references to produce the
paper-reference-paper cascade of two bipartite networks shown in
Figure \ref{f13} (a). Calculating the weights of this cascade using
matrix multiplication will produce the co-occurrence counts of
papers' links to references, bibliographic coupling strength
\cite{kessler63}, as was done in Equation (\ref{eq45}).

\begin{figure*}
\resizebox{0.9\textwidth}{!}{%
\includegraphics{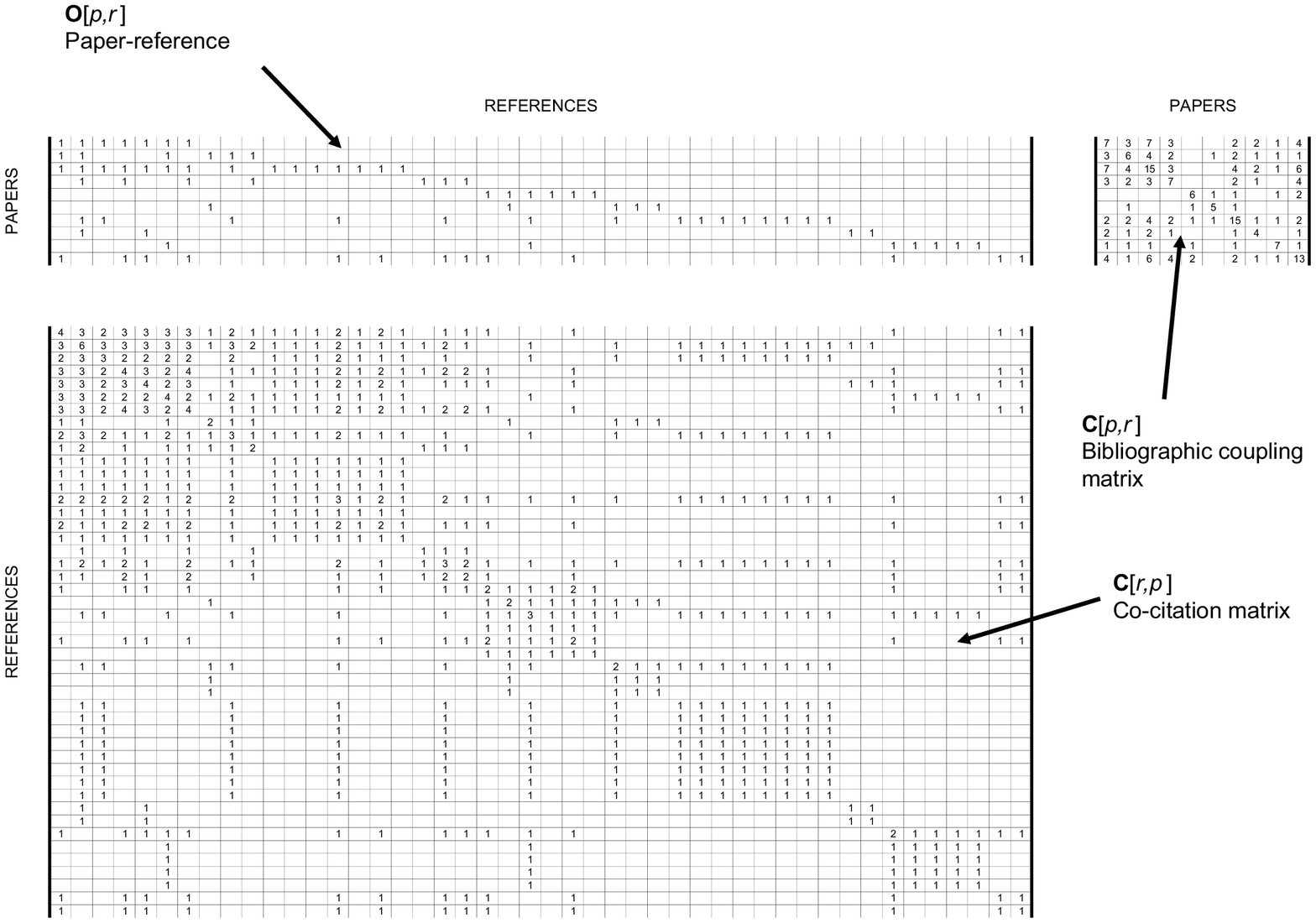}}%
\caption{Diagram showing that each occurrence matrix is associated
with a pair of co-occurrence matrices.  Upper left matrix is paper
to reference occurrence matrix $\mathbf{O}[p,r]$, below is reference
co-occurrence matrix relative to papers (co-citation matrix),
$\mathbf{C}[r,p]$.  Upper right matrix is paper co-occurrence matrix
relative to references (bibliographic coupling matrix),
$\mathbf{C}[p,r]$.\label{f14}}
\end{figure*}

The same network of Figure \ref{f3} can be mirrored on the papers to
produce the reference-paper-reference cascade of bipartite networks
shown in Figure \ref{f13}(b).  Calculating the link weights in this
network using matrix multiplication yields the co-occurrence counts
of references links to papers, co-citation strength \cite{small73}.
Note that each occurrence matrix has two co-occurrence matrices
associated with it. Figure \ref{f14} illustrates this for a sample
paper to reference occurrence matrix, $\mathbf{O}[p,r]$.  To the
right of $\mathbf{O}[p,r]$ is the square symmetric bibliographic
coupling matrix $\mathbf{C}[p,r]$, whose size is number of papers in
$\mathbf{O}[p,r]$. Similarly, below $\mathbf{O}[p,r]$ is the square
symmetric co-citation matrix, $\mathbf{C}[r,p]$ whose size is the
number of references in $\mathbf{O}[p,r]$.

Linguistic terms to paper networks, reference author to paper
networks and reference journal to paper networks  are weighted
networks. Because of this, it is not desirable to calculate their
co-occurrence matrices using matrix multiplication because the
resulting link weights cannot be interpreted.  Noting that
calculation of co-occurrence matrices is analogous to computing link
weights for a pair of cascaded bipartite networks, as was
demonstrated in Figure \ref{f13} and the discussion above, other
link weight functions can be used to find their co-occurrence
matrices. This can be done, for example, using the overlap function
of Section \ref{overlapsec}.

As an example, assume the paper to linguistic term matrix:
\begin{equation}
\mathbf{O}[p,t]=
 \left[\begin{array}{cccccc}
 8 & 9 & 5 & 3 & 1 & 0 \\
 5 & 4 & 9 & 2 & 0 & 1 \\
 0 & 0 & 2 & 6 & 5 & 4 \\
 1 & 1 & 0 & 5 & 2 & 5 \\
\end{array}\right]
\label{46} .\end{equation}
Using the overlap function, the
co-occurrence matrix of papers linked to terms is:
\begin{eqnarray}
\mathbf{C}[p,t]&=&OVL\Big(\mathbf{O}[p,t],\mathbf{O}[t,p]\Big) \nonumber \\
\nonumber \\
  &=&OVL\left(
 \left[\begin{array}{cccccc}
 8 & 9 & 5 & 3 & 1 & 0 \\
 5 & 4 & 9 & 2 & 0 & 1 \\
 0 & 0 & 2 & 6 & 5 & 4 \\
 1 & 1 & 0 & 5 & 2 & 5 \\
\end{array}\right],
 \left[\begin{array}{cccc}
 8 & 5 & 0 & 1 \\
 9 & 4 & 0 & 1 \\
 5 & 9 & 2 & 0 \\
 3 & 2 & 6 & 5 \\
 1 & 0 & 5 & 2 \\
 0 & 1 & 4 & 5 \\
\end{array}\right]\right)\nonumber \\
\nonumber \\
 &=& \left[\begin{array}{cccc}
 26 & 16 & 6 & 6 \\
 16 & 21 & 5 & 5 \\
 6 & 5 & 17 & 11 \\
 6 & 5 & 11 & 14 \\
\end{array}\right]
\label{eq47}
\end{eqnarray}

\section{Recursive matrix growth}
The recursive growth equations presented in this section are a
natural outgrowth of the proposed matrix-based  mathematical
treatment of collections of journal papers. They are useful for the
purpose of providing insight into the character of occurrence
distributions in the collections, as will be explained.

The basic record in a collection of journal papers is the paper. The
collection grows paper by paper in the temporal order of the
publication dates of the papers. When a new paper is added, it is
associated with the existing entities in the collection and
additionally, new entities, e.g., new paper authors or new
references, and new terms that enter into the collection.

This section will present a recursive model of the growth of both
occurrence and co-occurrence matrices as papers are added to the
collection.  The recursive model of matrix growth is found by
examination of matrix partitions in occurrence and co-occurrence
matrices as papers are added to the collection.

\begin{figure}
\resizebox{0.45\textwidth}{!}{%
\includegraphics{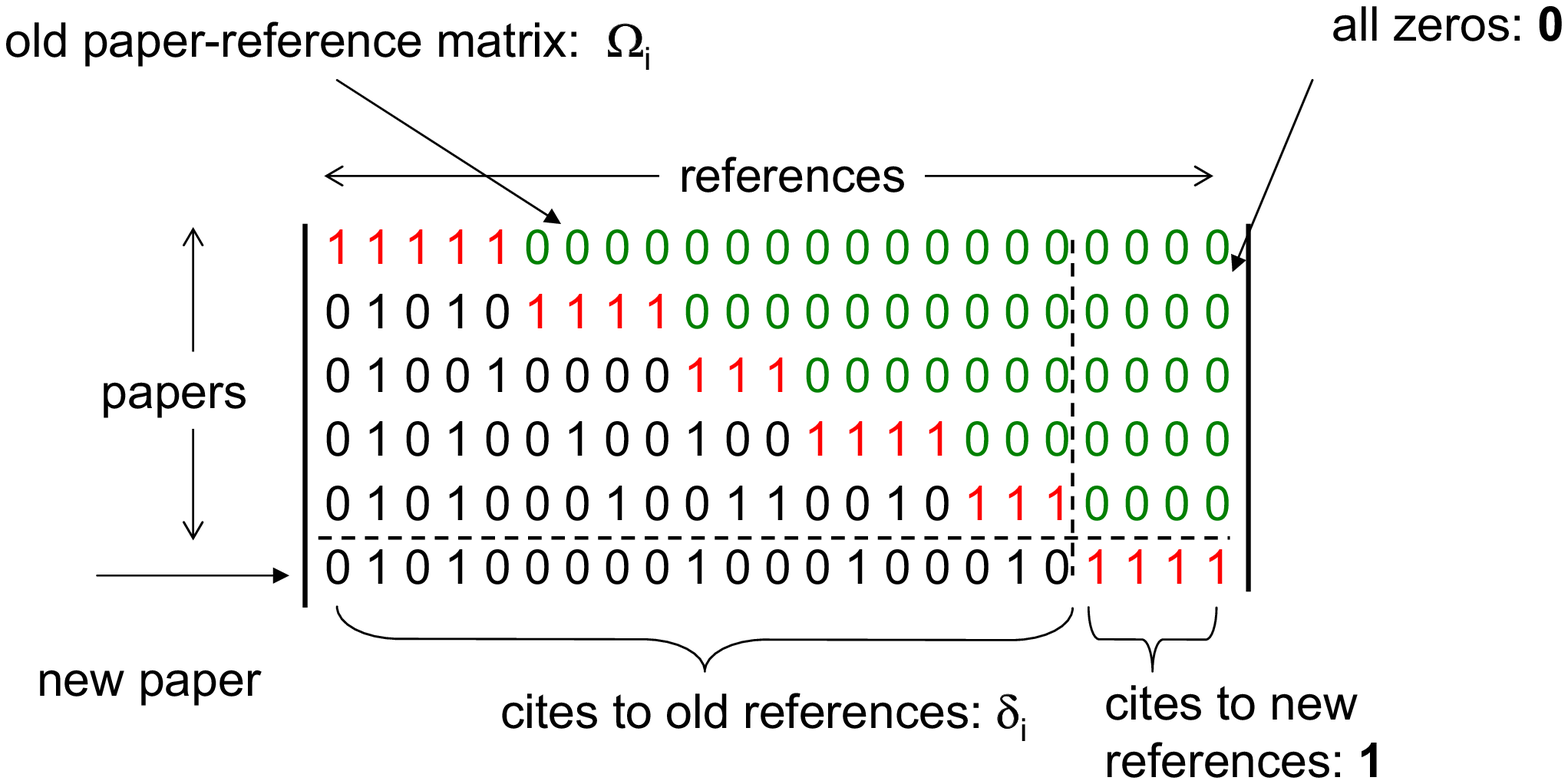}}%
\caption{Diagram of the structure of a paper to reference
matrix.\label{f23}}
\end{figure}

It is easiest to consider the growth of an example occurrence
matrix. For convenience, the paper-reference matrix will be studied.
The results can be easily extended to other occurrence matrices, for
example the paper to paper author matrix \cite{morris04b}. In the
matrix the rows correspond to papers and are ordered in the sequence
of publication of the papers to which they correspond. The columns
correspond to references and are ordered in the sequence in which
their corresponding references first appear. As shown in Figure
\ref{f23}, the matrix   contains a descending stair step sequence of
ones from its upper left corner diagonally to its lower right
corner. This sequence of ones corresponds to the initial appearance
of references as papers are added to the collection. Below this
diagonal sequence of ones is a roughly lower triangular region
sparsely populated with ones that correspond to citations to
existing references as each paper is added.  Above the diagonal
sequence of ones is a roughly upper triangular area of zeros.

\begin{figure}
\resizebox{0.45\textwidth}{!}{%
\includegraphics{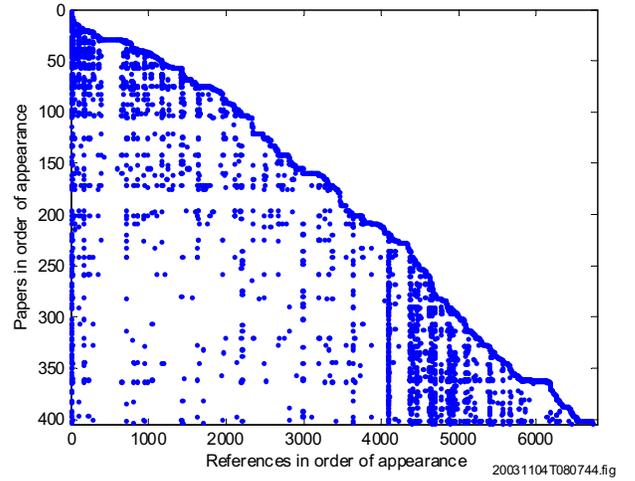}}%
\caption{Example of a typical paper to reference matrix.\label{f24}}
\end{figure}

Considering the collection of journal papers dynamically, the
collection grows from an initial paper by sequential addition of
papers in the order in which they were published.  In this sense the
paper-reference matrix $\bm{\Omega}$ grows dynamically one paper at
a time. Assume $i$ to be the number of papers, while $nr_i$ is the
number of references that have appeared in all papers up to and
including paper $i$. Assume $\bm{\Omega}_i$, whose size is $i$ by
$nr_i$, as the paper-reference matrix after the addition of paper
$i$, then consider the addition of paper $i+1$. A new row vector,
$i+1$, is added to  $\bm{\Omega}_i$. This vector is partitioned into
a 1 by $i$ vector $\bm{\delta}_i$ listing the paper's citations to
existing references, and $\mathbf{1}$, a 1 by $nr_{i+1}-nr_i$ vector
of ones occurring in new columns added for the new references that
have appeared in paper $i+1$. Figure \ref{f23} shows a pictorial
representation of this addition.  In the new columns, $\mathbf{0}$,
an $i$ by $nr_{i+1}-nr_i$ zero matrix appears. The recursive matrix
equation for growth of the paper-reference equation is:
\begin{equation}
\mathbf{\Omega}_{i+1}= \left[
\begin{array}{cc}
\mathbf{\Omega}_i & \mathbf{0} \\
\bm{\delta}_i & \mathbf{1} \\
\end{array}\right] \label{eq64}
.\end{equation}
Figure \ref{f24} shows a map of a typical
paper-reference matrix, where each dot shows the location of a one
in the matrix.

As papers are added to the collection, note that individual papers
collect no links after their initial appearance, while references
cumulate links (citations from newly appearing papers) as papers are
added. Entity-types that cumulate links in collections of papers
usually have a power-law frequency distribution relative to papers.
Three such power-law distributions are well-known: 1) papers per
paper author distribution  (Lotka's law) \cite{white89}, 2) papers
per paper journal distribution (Bradford's law) \cite{white89}, and
papers per reference distribution (reference power law)
\cite{naranan71}. Papers, which don't cumulate links, tend to have
exponential tailed distributions relative to other entity-types. Two
examples are authors per paper distribution (1-shifted Poisson)
\cite{morris04b}, and references per paper distribution (lognormal)
\cite{morris04a}.

\begin{figure}
\resizebox{0.45\textwidth}{!}{%
\includegraphics{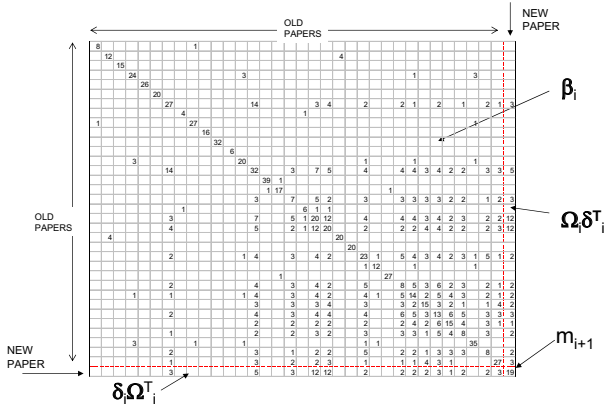}}%
\caption{Diagram of a bibliographic coupling matrix.\label{f25}}
\end{figure}

The bibliographic coupling matrix, which will be designated
$\bm{\beta}$, is a symmetric matrix that lists the bibliographic
coupling counts of all pairs of papers within the data collection.
The diagonal of $\bm{\beta}$ contains the counts of the number of
references cited in each paper. The bibliographic coupling matrix
can be obtained by multiplying the paper-reference matrix by its
transpose:
\begin{equation}
\bm{\beta} = \bm{\Omega}\cdot\bm{\Omega}^T \label{eq65}
.\end{equation}
The recursive growth equations for the bibliographic
coupling matrix can be derived by substituting (\ref{eq64}) into
(\ref{eq65}):
\begin{eqnarray}
\bm{\beta}_{i+1}&=& \bm{\Omega}_{i+1}\cdot\bm{\Omega}_{i+1}^T = \nonumber \\
\nonumber \\
 &=& \left[ \begin{array}{cc}
\bm{\Omega}_i\cdot\bm{\Omega}_i^T &
\bm{\Omega}_i\cdot\ \bm{\delta}_i^T \\
\bm{\delta}_i\cdot\bm{\Omega}_i^T &
\bm{\delta}_i\cdot\bm{\delta}_i^T + \bm{1}\cdot\bm{1}^T
\end{array}\right] \nonumber \\
\nonumber \\
 &=&\left[\begin{array}{cc}
\bm{\beta}_i & \bm{\Omega}_i\cdot\bm{\delta}_i^T \\
\bm{\delta}_i\cdot\bm{\Omega}_i^T & m_{i+1} \\
\end{array}\right]\label{eq66}
,\end{eqnarray} where $m_{i+1}$ is the number of references cited by
paper $i+1$. Figure \ref{f25} shows a pictorial representation of a
typical bibliographic coupling matrix with the partitions in
Equation (\ref{eq66}) identified.  It is easy to see from Equation
(\ref{eq66}) and Figure \ref{f25} that bibliographic coupling counts
between pairs of papers are static, and do not change as more papers
are added to the collection.

The co-citation matrix, designated as $\bm{\Gamma}$, is a symmetric
$nr$ by $nr$ matrix that lists the co-citation counts of all pairs
of references within the data collection.  The diagonal of
$\bm{\Gamma}$ contains the counts of the number of papers that cite
each reference.   The co-citation matrix can be obtained by
multiplying the transpose of the paper-reference matrix by itself:
\begin{equation}
\bm{\Gamma} = \bm{\Omega}^T\cdot\bm{\Omega} \label{eq67}
.\end{equation}
The recursive growth equations for the co-citation
matrix can be derived by substituting Equation (\ref{eq64}) into
Equation (\ref{eq67}):
\begin{eqnarray}
\bm{\Gamma}_{i+1}&=& \bm{\Omega}_{i+1}^T\cdot\bm{\Omega}_{i+1} \nonumber \\
\nonumber \\
 &=& \left[ \begin{array}{cc}
\bm{\Omega}_i^T\cdot\bm{\Omega}_i +
\bm{\delta}_i^T\cdot\bm{\delta}_i &
\bm{\delta}_i^T\cdot\ \bm{1} \\
\bm{1}^T\cdot\bm{\delta}_i^T & \bm{1}^T\cdot\bm{1}
\end{array}\right] \nonumber \\
\nonumber \\
 &=&\left[\begin{array}{cc}
\bm{\Gamma}_i + \bm{\delta}_i^T\cdot\bm{\delta}_i & \bm{\delta}_i^T\cdot\ \bm{1} \\
\bm{1}^T\cdot\bm{\delta}_i & \bm{1}^T\cdot\bm{1} \\
\end{array}\right]\label{eq68}
.\end{eqnarray}

\begin{figure}
\resizebox{0.45\textwidth}{!}{%
\includegraphics{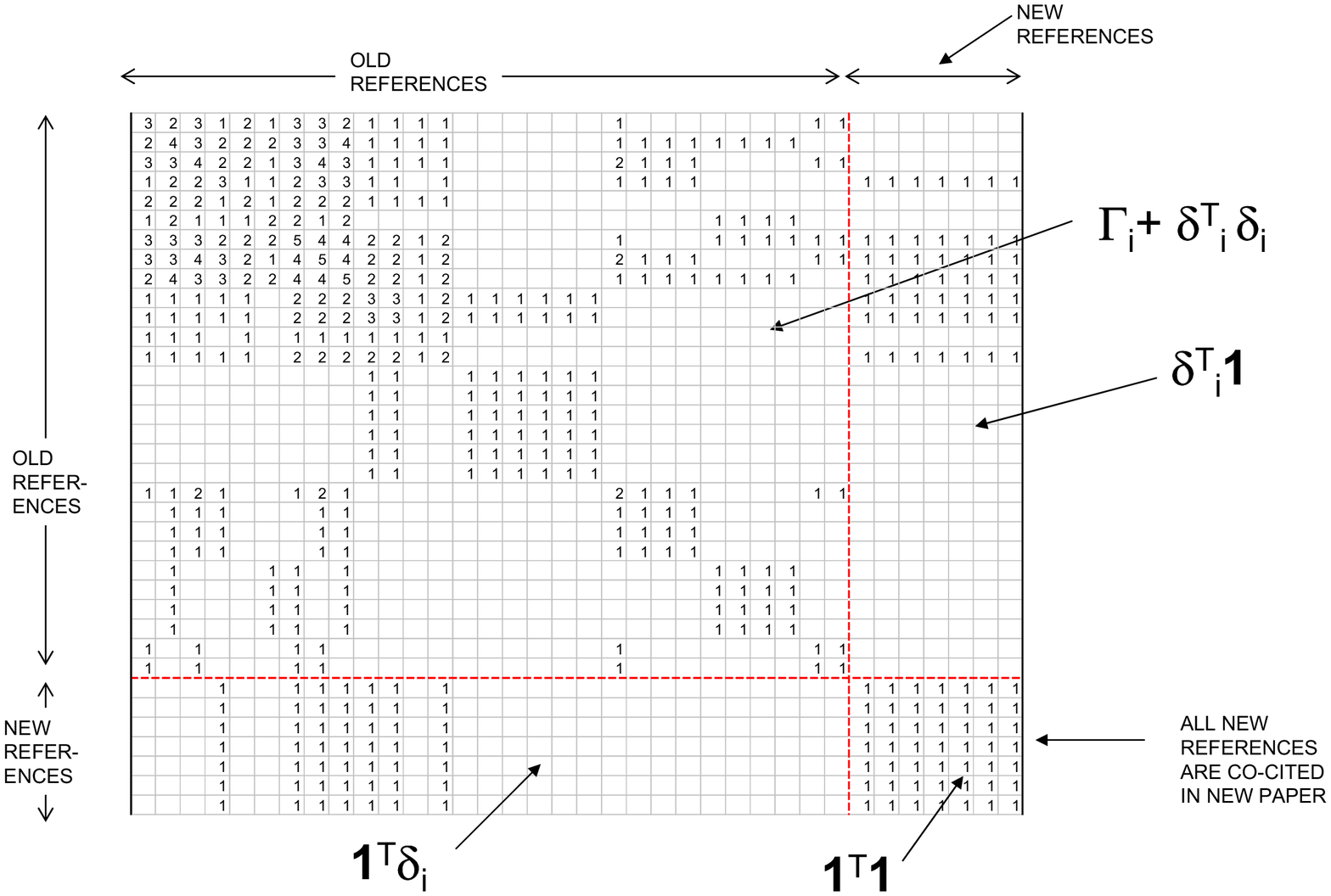}}%
\caption{Diagram of a co-citation matrix.\label{f26}}
\end{figure}

Figure \ref{f26} shows a pictorial representation of a typical
co-citation matrix with the partitions in Equation (\ref{eq68})
identified. It is easy to see that the co-citation count between two
references is not static, but can be increased with the addition of
each new paper to the collection.

\section{Example}
An illustrative example of the techniques outlined here uses a
collection of 902 papers on the topic of complex network theory.
This collection was gathered in 2003 by finding all papers that cite
key references in the specialty.  A detailed analysis of the paper
to reference network for this collection was presented by Morris
\cite{morris04a}, while analysis of the paper author to paper
network for this collection was presented by Goldstein, \emph{et
al}, \cite{goldstein04group} and Morris, \emph{et al},
\cite{morris04b}.

\begin{figure*}
\resizebox{1.0\textwidth}{!}{%
\includegraphics{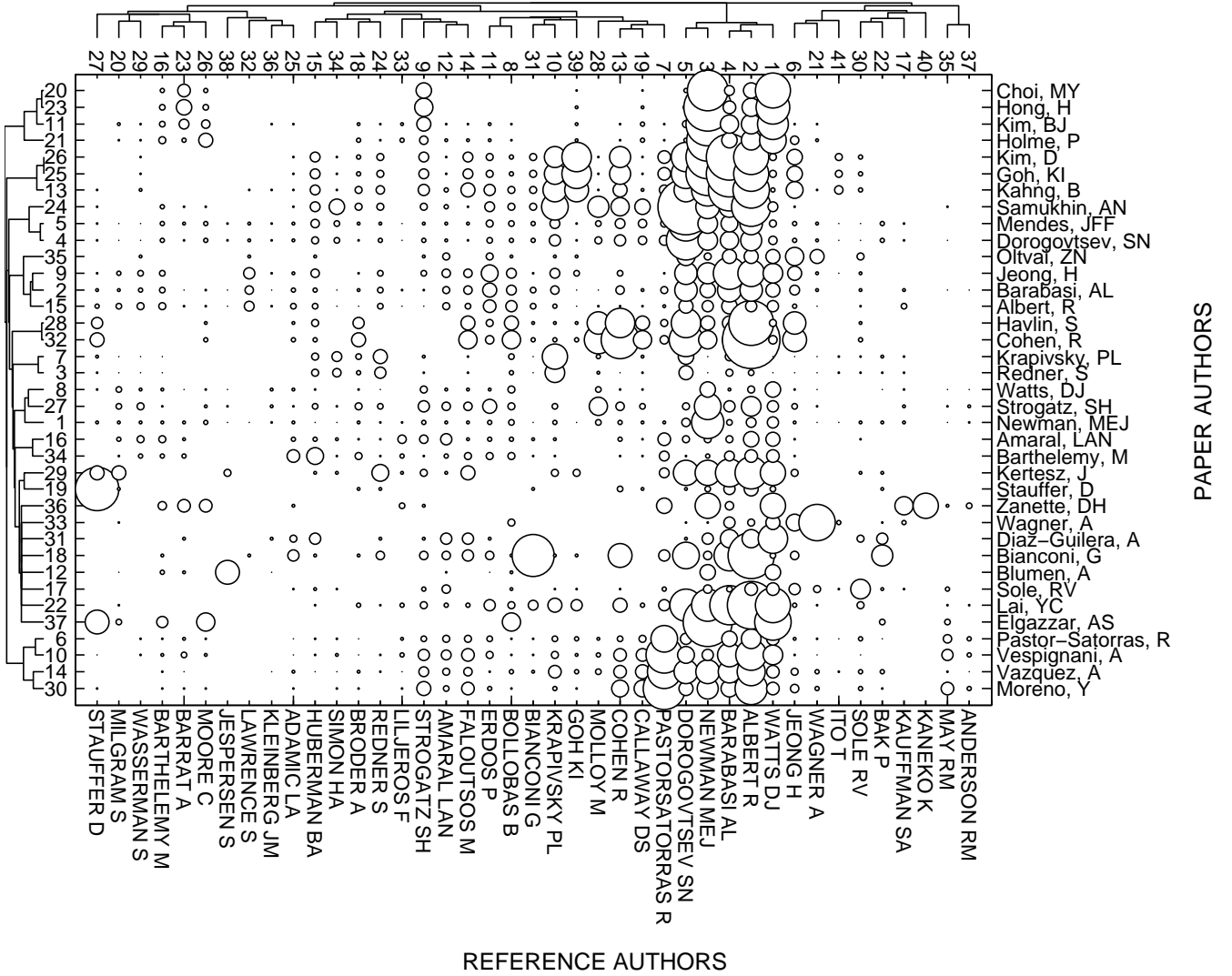}}%
\caption{Visualization of the occurrence matrix of a weighted paper
author to reference author network from a collection of papers from
the specialty of complex networks theory.\label{fap_ar}}
\end{figure*}

Figure \ref{fap_ar} shows a weighted occurrence matrix,
$\mathbf{O}[ap,ar]$, for the paper author to reference author
network from this collection. In this diagram, the paper authors are
rows, reference authors are columns, and the size of the circle at
position $(i,j)$ in the diagram is proportional to the link weight
from paper author $i$ to reference author $j$. In this case the link
weight is equal to the number of times that paper author $i$ cited
reference author $j$.

In order to visualize the structure of links in the network, the
rows and columns of the matrix have been arranged using a seriation
algorithm \cite{morris04optimal} and clustering dendrograms have
been added on the left and top of the figure
\cite{morris04crossmaps}. The figure is meant to show collaboration
groups of paper authors and their links to reference authors as
symbols of 'schools of thought' \cite{white97authors}. The
visualization technique of Figure \ref{fap_ar} is explained in
Morris and Yen \cite{morris04crossmaps}.

Only paper authors that authored 6 or more papers were visualized.
For clustering paper authors, the co-occurrence matrix of
co-authorship counts, $\mathbf{C}[ap,p]$, was calculated using
matrix multiplication:
$\mathbf{C}[ap,p]=\mathbf{O}[ap,p]\cdot\mathbf{O}[p,ap]$. These
co-authorship counts were converted to distances and a hierarchical
clustering routine was applied to produce the dendrogram on the left
of the figure. Groups of paper authors clustered this way can be
regarded as 'research teams.'

Only reference authors that were cited 50 or more times were
visualized. For clustering reference authors, the co-occurrence
matrix of co-citation counts, $\mathbf{C}[ar,p]$, was calculated
using the overlap function:
$\mathbf{C}[ar,p]=OVL(\mathbf{O}[ar,p],\mathbf{O}[p,ar])$. These
co-citation counts were converted to distances and a hierarchical
clustering routine was applied to produce the dendrogram at the top
of the figure. Groups of reference authors clustered this way can be
regarded as representing 'schools of thought.'

The paper author to reference author matrix, $\mathbf{O}[ap,ar]$,
was calculated using matrix multiplication
$\mathbf{O}[ap,ar]=\mathbf{O}[ap,p]\cdot\mathbf{O}[p,r]\cdot\mathbf{O}[r,ar]$.
The matrix clearly shows that dominant reference authors in the
specialty, who are cited by authors to represent key ideas in the
specialty, are heavily linked across all paper authors. Note that
there is evidence of correlation of groups of paper authors to
groups of reference authors. For example, paper authors Choi, Hong,
Kim and Holme are all heavily connected to reference authors Newman
and Watts, while paper authors Pastor-Satorras, Vespignani, Vazquez,
and Moreno are all heavily connected to reference authors
Pastor-Satorras and Albert.

This example illustrates the usefulness of the matrix-based
mathematical treatment of cascades of bipartite networks in
collection of journal papers. In the example, we have shown this
treatment can be used for construction of weighted unipartite
co-occurrence networks for clustering purposes: 1) paper authors
linked by co-authorship, and 2) reference authors linked by common
papers. Additionally, the method was used to calculate a weighted
bipartite network of paper authors to reference authors.

\section{Conclusion}

We have introduced several valuable methods that can be used to
apply complex networks theory to collections of journal papers:

\begin{itemize}
\item \textbf{The structural model of coupled bipartite networks for
collections of papers.} This is a novel model that allows analysis
of any bipartite network in the collection in a general,
standardized, manner. Further, it allows building a \emph{multiple
entity-type} growth model of this system of networks, a technique
not generally studied by complex networks researchers.
\item \textbf{The matrix-based method of calculating weighted bipartite
networks.} Using the general concept of link weight functions, we
have shown that this matrix-based technique can be applied to
cascades of unweighted bipartite networks using matrix
multiplicaiton. Additionally, the technique can be applied to
cascades of weighted bipartite networks using the overlap function
or the inverse Minkowski function.
\item \textbf{The calculation of weighted unipartite co-occurrence networks.}
 Considering co-occurrence networks as coupled bipartite networks
 made by mirroring around a bipartite partition, calculation of
 weighted co-occurrence networks uses the same matrix-based
 calculation method as weighted bipartite networks.
 \item \textbf{The construction of simple models of weighted matrix
 growth.} This structural model of coupled bipartite networks, when
 considered with unweighted bipartite growth models, such as the
 bipartite Yule model, yields a simple model of growth of weighted
 bipartite networks and weighted unipartite co-occurrence networks.
 Morris \cite{morris05a} has shown that simple bipartite Yule
 processes effectively simulate the statistics of bipartite and
 weighted unipartite  networks in collections of papers.
 \end{itemize}

The structural model and matrix-based techniques introduced here
provide a unified framework of all entities in networks of papers,
e.g., paper to author networks that are manifestations of social
collaboration processes, or paper to reference networks that are
manifestations of epistemological processes such as knowledge
accretion and exemplar knowledge in a specialty. Such networks are
often studied as decoupled processes despite their almost certain
interdependence. For example, note that the paper author to
reference author network example of Figure \ref{fap_ar} shows
correlations between groups of paper authors and groups of reference
authors. A realistic  model of processes in a research specialty
should be able to predict that such correlations will occur, but the
model must also predict the characteristics of the paper author to
paper network (such as Lotka's law), and simultaneously predict the
characteristics of the paper to reference network (such as the
reference power law.) All of these bipartite networks are
interdependent and those interdependencies cannot be modeled using
simple unipartite or bipartite growth models. The structural model
introduced here is a step toward modeling the complex
interdependencies in a research specialty.

Furthermore, and importantly, these techniques can be applied to
other report-based structures that can be expressed as collections
of entities.  For example, a collection of intelligence reports
about terrorist events can, after application of an entity
extraction program, be expressed as a collection of entities:
reports, place names, terrorist group leader names, terrorist group
names, government officials' names, and incident types. These
entities are linked in a coupled bipartite structure, similar to
Figure \ref{coupled} and analysis of those linkages could produce
useful information about networks of terrorists. So the structural
model introduced here may allow the study of other self-organizing
social organizations as well, through their manifestations in
collections of reports.

\section{Acknowledgements} We would like to thank Michel Goldstein,
now of Amazon.com, for many discussions and ideas that contributed
to this work over the last year.

\appendix*
\section{Example collection of journal papers}
\subsection{ISI tags\cite{isitags}}
The table below explains the tags used in the ISI source file given
in this appendix.

\begin{tabbing}
PT~\=Publication type\\
AU \>Author\\
TI \>Title\\
SO \>Source journal\\
ID \>Index terms\\
CR \>Cited reference\\
PY \>Published year\\
VL \>Volume\\
BP \>Beginning page\\
ER \>End of record\\
\end{tabbing}

\subsection{Source file}
Below, in ISI tagged file format, are listed four records comprising
a fictitious collection of papers on the fictional specialty of
\textit{improbability generation}:

\begin{footnotesize}
\texttt{
\begin{tabbing}
FN \= ISI \= Export Format\\
PT J\\
AU Beeblebrox, Z\\
TI Review of finite improbability generators\\
SO Bambleweeny Review \\
ID FINITE IMPROBABILITY; LIFE;  UNIVERSE\\
CR FORD P, 1996, J LIFE UNIV EVERY, V46, P111\\
   \>MOUSE B, 1997, REV FUT PHYS, V27, P76\\
   \>MOUSE B, 1998, BISTROMATH, V991, P342\\
PY 2003\\
VL 13\\
BP 844\\
ER\\
\\
PT J\\
AU \>Beeblebrox, Z\\
   \>Dent, A\\
   \>Prefect, F\\
TI Dentrassi hot tea: a scale free\\ \>\>brownian motion generator\\
SO Journal of Life, the Universe and Everything\\
ID FINITE IMPROBABILITY; ULTIMATE QUESTION;\\ \>\>  EVERYTHING, SCALE FREE NETWORKS\\
CR \>FORD P, 1996, J LIFE UNIV EVERY, V46, P111\\
   \>MOUSE B, 1998, BISTROMATH, V991, P342\\
   \>TRILLIAN A, 2000, SIRIAN CYBERN J, V82, P675\\
   \>TRILLIAN A, 2002, BISTROMATH, V995, P937\\
   \>BEEBLEBROX Z, 1994, REV FUT PHYS, V24, P923\\
PY 2003\\
VL 56\\
BP 738\\
ER\\
\\
PT J\\
AU Prefect, F\\
TI Application of infinite improbability to\\ \>\> spacecraft propulsion\\
SO Bistromathematica\\
ID INFINITE IMPROBABILITY; LIFE;  UNIVERSE; EVERYTHING\\
CR FORD P, 1996, J LIFE UNIV EVERY, V46, P111\\
   \>MOUSE B, 1998, BISTROMATH, V991, P342\\
   \>TRILLIAN A, 2002, BISTROMATH, V995, P937\\
   \>BEEBLEBROX Z, 2003, BAMBLEWEENY REV, V13, P844\\
   \>BEEBLEBROX Z, 1989, PRINCIPLES OF IMPROBAPHYSICS\\
PY 2004\\
VL 997\\
BP 938\\
ER\\
\\
PT J\\
AU Prefect, F\\
TI Power laws in infinite improbability networks\\
SO Proc of Vogonian Academy of Science\\
ID INFINITE IMPROBABILITY; ULTIMATE QUESTION;\\ \>\>EVERYTHING; SCALE FREE NETWORKS\\
CR FORD P, 1996, J LIFE UNIV EVERY, V46, P111\\
   \>MOUSE B, 1997, REV FUT PHYS, V27, P76\\
   \>TRILLIAN A, 2000, SIRIAN CYBERN J, V82, P675\\
   \>BEEBLEBROX Z, 2003, BAMBLEWEENY REV, V13, P844\\
   \>SLARTIBARTFAST B, 2001, GALACT J PHYS, V887, P2846 \\
   \>ZARNIWOOP N, 1978, MEGADODO MAG, V564, P23\\
PY 2004\\
VL 83\\
BP 944\\
ER\\
\\
EF\\
\end{tabbing}
} 
\end{footnotesize}

\subsection{Extracted entities} The table below lists the entities
extracted from the collection of papers above.

\begin{footnotesize}
\texttt{
\begin{tabbing}
\textbf{Papers} (identified by title)\\
$p_1$:~~\= Rev\=iew of finite\dots \\
$p_2$:  \> Dentrassi hot tea: a scale free\dots\\
$p_3$:  \> Application of infinite\dots \\
$p_4$:  \> Power laws in infinite\dots \\
\\
\textbf{Paper authors} \\
$ap_1$:  \> Beeblebrox, Z.\\
$ap_2$:  \> Dent, A.\\
$ap_3$:  \> Prefect, F \\
\\
\textbf{Paper journals} \\
$jp_1$:  \> Bambleweeny Review\\
$jp_2$:  \> Journal of Life, the Universe and Everything\\
$jp_3$:  \> Bistromathematica \\
$jp_4$:  \> Proc of Vogonian Academy of Science \\
\\
\textbf{Index terms} \\
$t_1$:  \> FINITE IMPROBABILITY\\
$t_2$:  \> LIFE\\
$t_3$:  \> UNIVERSE \\
$t_4$:  \> ULTIMATE QUESTION \\
$t_5$:  \> EVERYTHING \\
$t_6$:  \> SCALE FREE NETWORKS \\
$t_7$:  \> INFINITE IMPROBABILITY \\
\\
\textbf{References} \\
$r_1$:  \> FORD P, 1996, J LIFE UNIV EVERY, V46, P111\\
$r_2$:  \> MOUSE B, 1997, REV FUT PHYS, V27, P76\\
$r_3$:  \> MOUSE B, 1998, BISTROMATH, V991, P342\\
$r_4$:  \> TRILLIAN A, 2000, SIRIAN CYBERN J, V82, P675\\
$r_5$:  \> TRILLIAN A, 2002, BISTROMATH, V995, P937\\
$r_6$:  \> BEEBLEBROX Z, 1994, REV FUT PHYS, V24, P923\\
$r_7$:  \> BEEBLEBROX Z, 2003, BAMBLEWEENY REV, V13, P844\\
$r_8$:  \> BEEBLEBROX Z, 1989, PRINCIPLES OF IMPROBAPHYSICS\\
$r_9$:  \> SLARTIBARTFAST B, 2001, GALACT J PHYS, V887, P2846\\
$r_{10}$: \> ZARNIWOOP N, 1978, MEGADODO MAG, V564, P23\\
\\
\textbf{Reference authors} \\
$ar_1$:  \> FORD P\\
$ar_2$:  \> MOUSE B\\
$ar_3$:  \> TRILLIAN A\\
$ar_4$:  \> BEEBLEBROX Z\\
$ar_5$:  \> SLARTIBARTFAST B\\
$ar_{6}$: \> ZARNIWOOP N\\
\\
\textbf{Reference journals} \\
$jr_1$:  \> J LIFE UNIV EVERY \\
$jr_2$:  \> REV FUT PHYS \\
$jr_3$:  \> BISTROMATH \\
$jr_4$:  \> SIRIAN CYBERN J\\
$jr_5$:  \> BAMBLEWEENY REV\\
$jr_6$:  \> GALACT J PHYS \\
$jr_7$: \> MEGADODO MAG \\
\end{tabbing}
}
\end{footnotesize}

\subsection{Occurrence matrices}
Below are the occurrence matrices for the the direct bipartite
networks in the collection of papers above.

Paper to reference network:

\begin{equation}
\mathbf{O}[p,r]=
 \left[
\begin{array}{cccccccccc}
 1 & 1 & 1 & 0 & 0 & 0 & 0 & 0 & 0 & 0 \\
 1 & 0 & 1 & 1 & 1 & 1 & 0 & 0 & 0 & 0 \\
 1 & 0 & 1 & 0 & 1 & 0 & 1 & 1 & 0 & 0 \\
 1 & 1 & 0 & 1 & 0 & 0 & 1 & 0 & 1 & 1 \\
\end{array}
\right] \label{opr}
\end{equation}

Paper to paper author network:

\begin{equation}
\mathbf{O}[p,ap]=
 \left[
\begin{array}{ccc}
 1 & 0 & 0 \\
 1 & 1 & 1 \\
 0 & 0 & 1 \\
 0 & 0 & 1 \\
\end{array}
\right]\label{opap}
\end{equation}

Paper to paper journal network:

\begin{equation}
\mathbf{O}[p,jp]=
 \left[
\begin{array}{cccc}
 1 & 0 & 0 & 0 \\
 0 & 1 & 0 & 0 \\
 0 & 0 & 1 & 0 \\
 0 & 0 & 0 & 1 \\
\end{array}
\right]\label{opjp}
\end{equation}

Paper to terms network:

\begin{equation}
\mathbf{O}[p,t]=
 \left[
\begin{array}{ccccccc}
 1 & 1 & 1 & 0 & 0 & 0 & 0\\
 1 & 0 & 0 & 1 & 1 & 1 & 0\\
 0 & 1 & 1 & 1 & 0 & 0 & 1\\
 0 & 0 & 0 & 1 & 1 & 1 & 1\\
\end{array}
\right]\label{opt}
\end{equation}

Reference to reference author network:

\begin{equation}
\mathbf{O}[r,ar]=
 \left[\begin{array}{cccccc}
 1 & 0 & 0 & 0 & 0 & 0 \\
 0 & 1 & 0 & 0 & 0 & 0 \\
 0 & 1 & 0 & 0 & 0 & 0 \\
 0 & 0 & 1 & 0 & 0 & 0 \\
 0 & 0 & 1 & 0 & 0 & 0 \\
 0 & 0 & 0 & 1 & 0 & 0 \\
 0 & 0 & 0 & 1 & 0 & 0 \\
 0 & 0 & 0 & 1 & 0 & 0 \\
 0 & 0 & 0 & 0 & 1 & 0 \\
 0 & 0 & 0 & 0 & 0 & 1 \\
\end{array}\right]
\label{orar}
\end{equation}
\vspace{2in}

Reference to reference journal network:

\begin{equation}
\mathbf{O}[r,jr]=
 \left[\begin{array}{ccccccc}
 1 & 0 & 0 & 0 & 0 & 0 & 0\\
 0 & 1 & 0 & 0 & 0 & 0 & 0\\
 0 & 0 & 1 & 0 & 0 & 0 & 0\\
 0 & 0 & 0 & 1 & 0 & 0 & 0\\
 0 & 0 & 1 & 0 & 0 & 0 & 0\\
 0 & 1 & 0 & 0 & 0 & 0 & 0\\
 0 & 0 & 0 & 0 & 1 & 0 & 0\\
 0 & 0 & 0 & 0 & 0 & 0 & 0\\
 0 & 0 & 0 & 0 & 0 & 1 & 0\\
 0 & 0 & 0 & 0 & 0 & 0 & 1\\
\end{array}\right]
\label{orjr}
\end{equation}

Note that in some cases the paper to terms matrix may weighted when
working with abstract or title terms rather than index terms.

\bibliography{weight}

\end{document}